\documentclass[doublecol]{epl2}

\title{Formation of Modularity in a Model of Evolving Networks}
\shorttitle{Formation of Modularity in a Model of Evolving Networks} 

\author{Menghui Li\inst{1,2} \and Shuguang Guan\inst{3} \and Choy-Heng Lai \inst{2,4}}
\shortauthor{M. H. Li \etal}
 \institute{
  \inst{1} Temasek Laboratories, National University of
Singapore, 117508, Singapore\\
  \inst{2} Beijing-Hong
Kong-Singapore Joint Centre for Nonlinear \& Complex Systems
(Singapore), National University of Singapore, Kent Ridge, 119260,
Singapore\\
   \inst{3} Institute of Theoretical Physics and Department of
Physics, East China Normal University, Shanghai, 200062, P. R.
China\\
   \inst{4} Department
of Physics, National University of Singapore, 117542, Singapore
 }

 \pacs{89.75.Hc}{Networks and genealogical trees }
 \pacs{05.45.Xt}{Synchronization; coupled oscillators}
 \pacs{05.65.+b}{Self-organized systems}

\abstract{ Modularity structures are common in various social and
biological networks. However, its dynamical origin remains an open
question. In this work, we set up a dynamical model describing the
evolution of a social network. Based on the observations of real
social networks, we introduced a link-creating/deleting strategy
according to the local dynamics in the model. Thus the coevolution
of dynamics and topology naturally determines the network
properties. It is found that for a small coupling strength, the
networked system cannot reach any synchronization and the network
topology is homogeneous. Interestingly, when the coupling strength
is large enough, the networked system spontaneously forms
communities with different dynamical states. Meanwhile, the network
topology becomes heterogeneous with modular structures. It is
further shown that in a certain parameter regime, both the degree
and the community size in the formed network follow a power-law
distribution, and the networks are found to be assortative. These
results are consistent with the characteristics of many empirical
networks, and are helpful to understand the mechanism of formation
of modularity in complex networks.}

\begin{document}
 \maketitle

Real-world complex networks usually have certain universal
properties, such as small-world, scale-free, and modularity
\cite{Modular:empirical,Modular:review,CN:REV,CN:Mobile}. Scale-free
means that the degree of a network follows a power-law distribution,
and the modularity refers to the fact that networks typically
consist of communities or clusters in which the nodes are more
highly connected to each other than that to the rest of the network
\cite{Modular:empirical}. In the past decade, the investigation on
networked systems is mainly focused on two aspects. On the one hand,
the collective behaviors of networked oscillators have been
extensively studied. Most of these works are carried out on static
networks, aiming at revealing how network topology affects dynamics
\cite{Oscillator}. On the other hand, many evolving network models
have been proposed from the topological perspective, aiming at
constructing networks with power-law distribution of degree
\cite{CN:REV} and modular structure
\cite{Modularity:topology1,Modularity:topology2}. However, from
functional perspective, taking biological networks for instance,
modularity is generally believed to correspond to certain functional
groups, where nodes within the same group may share similar
characteristics \cite{functional}. This implies that in principle
the network topology and dynamics are strongly dependent on each
other. In fact, any formed network structure and dynamical pattern
are actually the result of coevolution of both network dynamics and
topology \cite{adaptive:REV}. For example, in various biological and
social networked systems, such as the email network
\cite{empirical:email} and the mobile communication network
\cite{CN:Mobile}, individuals are more likely to interact with
others ``similar" to themselves \cite{interactsimialr}, which
usually leads to networks consisting of communities driven by shared
activities, attributes, affiliations, and so on.

The formation mechanism of modularity and scale-free property is
crucial to the understanding of structural and functional/dynamical
properties of complex networks. Recently increasing attentions have
been paid to the adaptive coevolutionary networks
\cite{adaptive:REV,adaptive:rewir,adaptive:group,adaptive:SYN,adaptive:weight,
adaptive:community,adaptive:competitive,adaptive:opinion,adaptive:twogroup,adaptive:group-based,coevolution,adaptive:dilemma,adaptive:PD,stability}.
For example, Ref. \cite{adaptive:weight} investigated the
interaction between link weight and dynamical states on networked
phase oscillators. However, so far, it has not been well understood
how modular structure and power-law degree distribution concurrently
emerge in an evolving network. Motivated by the above idea, in the
present work, we set up a model of dynamical network whose nodes are
represented by phase oscillators. Basically, the model is an
evolving network which grows from a few nodes at the very beginning.
The most important characteristic of the model is that it describes
the interplay between the topological structure and the dynamics on
the network. On the one hand, the node dynamics are coupled to each
other according to the network topology; on the other hand, the
connections among nodes can be created or deleted according to the
local dynamical states. Our particular interest is focused on what
types of network structures can be formed as a result of the
coevolution of network dynamics and topology. Mainly, our study
presented the following results: (i) The communities, within which
the nodes have similar dynamical states and the connections are
denser than outside, can be naturally formed during the network
evolution. (ii) In a certain parameter regime, the degree
distribution and the distribution of community size follow power
law. (iii) The networks turn out to be assortative, i.e., the nodes
with high
 degree tend to connect to other nodes with high degree. In addition,
the average clustering coefficient of the network is generally high.
These properties exhibited by the model concurrently emerge as the
network evolves, and are consistent with the observations in many
realistic networks
\cite{Modular:empirical,Modular:review,CN:REV,CN:Mobile,empirical:email,communitysize,correlation,clustering3,k-clique}.

In real networked dynamical system, the individual dynamics is
generally complicated and different in principle, which is beyond
our capability of mathematical treatment so far. Therefore, the
local dynamics on networks is usually simplified as continuous
oscillator, or even discrete map in theoretical study
\cite{adaptive:REV,adaptive:rewir,adaptive:group,adaptive:SYN,adaptive:weight,adaptive:community,adaptive:competitive,adaptive:opinion,adaptive:twogroup,adaptive:group-based,adaptive:dilemma,adaptive:PD}.
In our model, the node dynamics are represented by phase
oscillators, which are coupled as in the following dynamical
equations:
\begin{equation}
\dot{\theta}_m = \omega_m + \frac{\gamma}{k_m} \sum_{n=1}^{N}a_{mn}
\sin(\theta_n - \theta_m+ \phi_{mn}). \label{model}
\end{equation}
Here, the dynamical state $\theta$ describes the attributes of
oscillator (node). $m,n=1,2,\ldots,N$ are the oscillator (node)
indices, and $\gamma$ is the uniform coupling strength. $A=\{ a_{mn}
\}$ is the adjacency matrix, where $a_{mn}=1$ if nodes $m$ and $n$
are connected, and $a_{mn}=0$ otherwise. $\{\omega_m\}$ are the
intrinsic frequencies of oscillators, and $k_m$ is the degree of
oscillator $m$.  We noticed that in many social networks individuals
tend to contact with others with similar attributes, i.e.,
similarity breeds connection \cite{interactsimialr}. Furthermore, if
two individuals are in the same environment, they are also more
likely to make friends with each other. For example, when two
individuals study in one class, or they work in the same company,
they are more likely to contact each other \cite{empirical:email}.
Considering the influence of environment, in our model we
particularly introduced an extra phase coupling term $\phi_{mn}$ as
\begin{equation}
\phi_{mn}=(\psi_m-\psi_n)\  mod \  (\pi).
\end{equation}
Here $\psi_m$ is the average phase of the local order parameter
which is defined as
\begin{equation}
r_{m}e^{i\psi_m}=\frac{1}{k_m}\sum_{n=1}^{N}a_{mn} e^{i\theta_n}.
\label{localparameter}
\end{equation}

In an adaptive networked system, the network topology usually
changes according to the dynamical interaction among nodes during
the evolution. Our model is basically a growing network which starts
from a few seed nodes at the very beginning. Particularly, the
evolution rule of the network incorporates two main manipulations:
one is the node-adding, and the other is link-adjusting, including
adding and removing links. It has been shown that in empirical
networks \cite{empirical:email,GrowthFlickr}, most new connections
are likely to build up between one node and its second neighbors,
i.e., the neighbor's neighbor. Based on this idea, in our model we
proposed a link-adjusting strategy according to the local dynamical
states. Depending on the local order parameter $r_m$, an active
individual has two options to adjust its link. One option is that if
his neighbors do not reach a consensus, the individual is free and
can link to any other individuals. Otherwise, if his neighbors reach
a consensus, he is ``frozen" in his neighborhood and can only make
new link to his second neighbors. Moreover, existing links may be
removed from the existing network for various reasons
\cite{tieeliminate}. This effect has also been considered in our
model.

\begin{figure}[tbp]
\begin{center}
\includegraphics[width=0.9\linewidth]{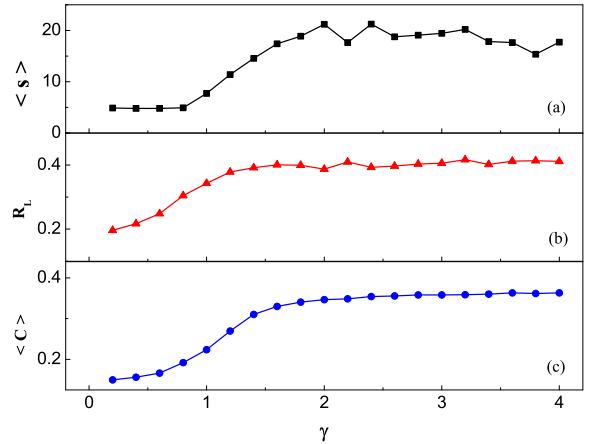}
\caption{(Color online) Characterization of the properties of
networks as a function of uniform coupling strength $\gamma$. (a)
The average cluster size $\langle s\rangle$ for 4-clique. (b) The
relative largest community size $R_L$ for  4-clique. (c) The average
clustering coefficient $\langle C\rangle$. Results are averaged over
$100$ realizations. The parameters are $N_0=5$, $T_r=10.0$,
$T_a=100.0$, $p=0.5$, and the final network size $N=500$. }
\label{fig1}
\end{center}
\end{figure}

Specifically, the dynamics and topology of the networked system
coevolve according to the following rules.

\begin{enumerate}
\item At the very beginning, the network consists of $N_0$ isolated
oscillators, whose dynamical states evolve according to Eq. (\ref{model}).
 The initial states of oscillators are randomly selected from $[-\pi,\pi]$,
 and their intrinsic frequencies are chosen from the distribution of $g(\omega)=0.75(1-\omega^2)$, where $\omega \in (-1,1)$.
 Eq. (\ref{model}) is integrated by the fourth order Runge-Kutta method with the time step  $\Delta t
 =0.01$.

\item  At every $T_a$ time ($T_a=m\Delta t, m\in\mathcal{N}$), one new node is added into the network by connecting it to an arbitrarily selected node in the existing network.

\item At every $T_r$ time ($T_r=n\Delta t, n\in\mathcal{N}$), one node $m$, randomly selected from the
    existing network, will be activated to adjust his links. Suppose the local order parameter of node $m$ is $r_m$, which characterizes the local
    coherence around node $m$. Then with total probability $r_m$,
    the node $m$ first chooses one of its nearest neighbors $k$ according to the following probability partition function
\begin{equation}
\prod^{select} =f(m,k)= \frac{L_{mk}(\Delta
\theta_{mk})}{\sum_{j\in\partial_m}L_{mj}(\Delta \theta_{mj})},
\label{selecting}
\end{equation}
where $\partial_m$ denotes the set of the nearest neighbors of the
node $m$. $L_{mk}(\Delta \theta_{mk})=[1+\cos(\Delta
\theta_{mk})]/2$ is the similarity distance
\cite{model:local,measure}. According to the definition, the larger
the similarity distance is, the closer the dynamical states between
these two connecting oscillators are. Then the node $k$ introduces
one of its nearest neighbor $i$ to the node $m$ according to
probability partition function
\begin{equation}
\prod^{add}_{m\rightarrow i} =f(k,i)_{i\in \partial_k, i\not\in
\partial_m}. \label{adding}
\end{equation}
In this way, a new link between $m$ and one of its second neighbor
$i$ is established. Meanwhile, with the total probability $1-r_m$,
the node $m$ connects to an existing node in the network according
to probability partition function $f(m,j)_{j\not\in\partial_m}$.

\item Parallel to the above step, the node $m$ will remove one of its
existing connections with the total probability $p$ according to the
following probability partition function
\begin{equation}
\prod^{cut} = \frac{1/L_{mk}(\Delta
\theta_{mk})}{\sum_{j\in\partial_m}1/L_{mj}(\Delta \theta_{mj})}.
\label{cutting}
\end{equation}
This implies that the probability removing the link between $m$ and
$k$ is inversely proportional to their similarity distance.
\end{enumerate}
Following the above network evolution rules, actually there are two
time scales respectively characterized by $T_a$ and $T_r$ in our
model. In our simulations, we consider the situation that adding new
node is slower than adding/removing links, i.e., $T_a$ is chosen to
be larger than $T_r$. In the rare cases where these rules would lead
to multiple links we do not allow them to be formed.

\begin{figure}[tbp]
\begin{center}
\includegraphics[width=\linewidth]{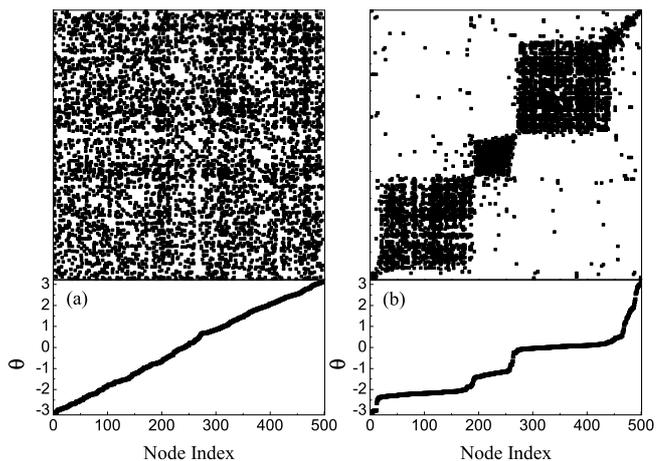}
\caption{ Typical examples of  the final networks: the adjacent
matrix (top) and the final phase states of oscillators (bottom). (a)
$\gamma=0.2$. Approximately, the network is homogeneous and the
phases of oscillators follow a uniform distribution. (b) $\gamma=6$.
Communities is formed, and the oscillators within the same community
have close phase states. The indices of the oscillators have been
rearranged according to the phase. Other parameters are the same as
those in Fig. \ref{fig1}. } \label{fig2}
\end{center}
\end{figure}

\begin{figure*}[tbp]
\begin{center}
\includegraphics[width=0.7\linewidth]{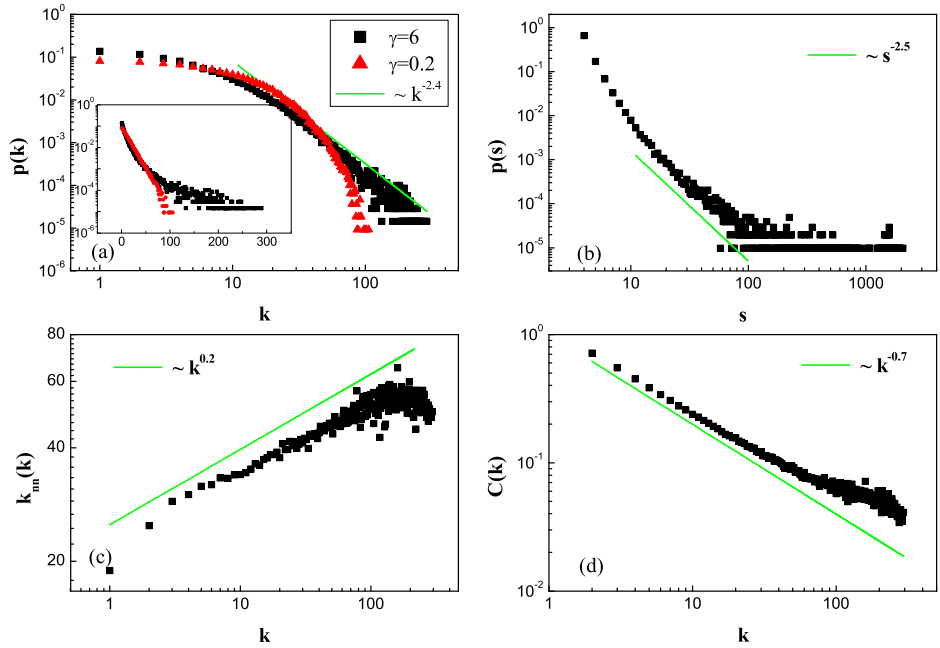}
\caption{(Color online) The topological properties of final networks
of $N=10000$. (a) The distribution of degree. Note that the inset
plot is on log-normal coordinate. (b) The distribution of community
size for 4-clique, where $\gamma=6$. (c) The average nearest
neighbor degree $k_{nn}(k)$ as a function of degree $k$, where
$\gamma=6$. (d) The average clustering coefficient $C(k)$ as a
function of degree $k$, where $\gamma=6$. The straight line is a
guide to the eye. Other parameters are the same as those in Fig.
\ref{fig1}. } \label{fig3}
\end{center}
\end{figure*}

According to the above rules, given the parameters, the formed
network will approximately have a constant average degree
\begin{equation}\label{degree-distribution}
\langle k \rangle\approx 2[1+\frac{T_a}{T_r}(1-p)],
\end{equation}
which is independent on the network size. However, the degree
distribution can be rather heterogeneous. In order to investigate
the possible communities of the formed network, we employed the
$k$-clique percolation method, in which communities can be defined
in terms of adjacent cliques \cite{k-clique}. This method has
advantage to avoid the problem of resolution limit in many other
community identification methods
\cite{Modular:review,resolutionlimit}. Specifically, we calculated
the following quantities to characterize the community structure in
network: the average cluster size $\langle s\rangle$ (except the
largest one), the relative largest cluster size $R_L$ and the
average clustering coefficient $\langle C\rangle$. The average
community size $\langle s\rangle$ is defined as
\begin{equation}
\langle s\rangle
=\frac{\sum_sn_ss^2}{\sum_sn_ss},\label{secondmoment}
\end{equation}
where $s$ is the size of cluster and $n_s$ is the number of $s$-size
cluster. The sums run over all possible values of $s$ but the
largest cluster. The relative largest cluster size $R_L$ is defined
as
\begin{equation}
R_L =\frac{s_L}{N},\label{RL}
\end{equation}
where $s_L$ is the size of the largest cluster, and $N$ is the size
of system. In our model, the active node can connect to one of its
second neighbors with a certain probability, which generates at
least one triangle. Therefore, we focused on 4-clique and 5-clique
in our computation. Actually, our numerical experiments show that
they give qualitatively consistent results.

Now we report the main results of our numerical simulations. We
first study how the network evolves with the increase of the
coupling strength. Figure \ref{fig1} shows various properties of the
final networks with respect to the coupling strength $\gamma$. It is
found that when $\gamma\rightarrow 0$, both $\langle s \rangle$ and
$R_L$ are very small. This means that the network structure is
basically homogeneous and no obvious clusters are formed at this
stage. However, the network structure significantly changes when
$\gamma$ is large enough. For instance, when $\gamma>2 $ , both
$\langle s \rangle$ and $R_L$ are significantly large indicating
that the network structure is characterized by limited number of
communities. This can be further verified by the average clustering
coefficient $\langle C \rangle$ as shown in Fig. \ref{fig1}(c). A
careful examination of the dynamical states on the network shows
that when the coupling strength $\gamma$ is small, the system can
reach neither synchronization nor clustering. In this case, the
links are almost randomly generated, so the formed network is a
random one. There are no obvious communities as shown in Fig.
\ref{fig2}(a). On the contrary, when $\gamma$ is sufficiently large,
clustering occurs in the networked system with the growth of the
network size. Physically, there are two factors affecting the
formation of dynamical groups. On the one hand, according to the
network-growing rule, a node is more likely to connect to its second
neighbors with similar states. This will gradually generate a core
of oscillators which are partially synchronized. On the other hand,
if two oscillators are in two different dynamical groups, their
``environment" are different, i.e., $\phi_{mn}$ between them is
large \cite{linkfunction}. Since large $\phi_{mn}$ is not in favor
of synchronization \cite{Phaselag}, these two nodes have less chance
to build a connection between them. In addition, once the phase
difference between two oscillators is large, it is likely that the
existing link between them will be disconnected, while new link is
hardly generated between them. Therefore, once a dynamical group,
i.e., a core of oscillators with similar dynamical states and
similar frequencies, is generated, the network-growing mechanism of
the present model will enhance the formation of dynamical groups, as
well as the formation of dense connections inside the group.
Numerically, it is observed that after a long time evolution, the
oscillators self-organize into many communities, both dynamically
and topologically as shown in Fig. \ref{fig2}(b). In our model, the
coupling strength $\gamma$ stands for the magnitude of strength of
interaction among different nodes in the network. Therefore, the
above results emphasize one important fact that during the evolution
of networked system, the network structure can be significantly
affected by the interaction strength among nodes. Conversely, if we
want to correctly analyze the formation mechanism of modularity in a
network, we have to pay more attention on the specific dynamical
processes on it.

\begin{figure*}[tbp]
\begin{center}
\includegraphics[width=\linewidth]{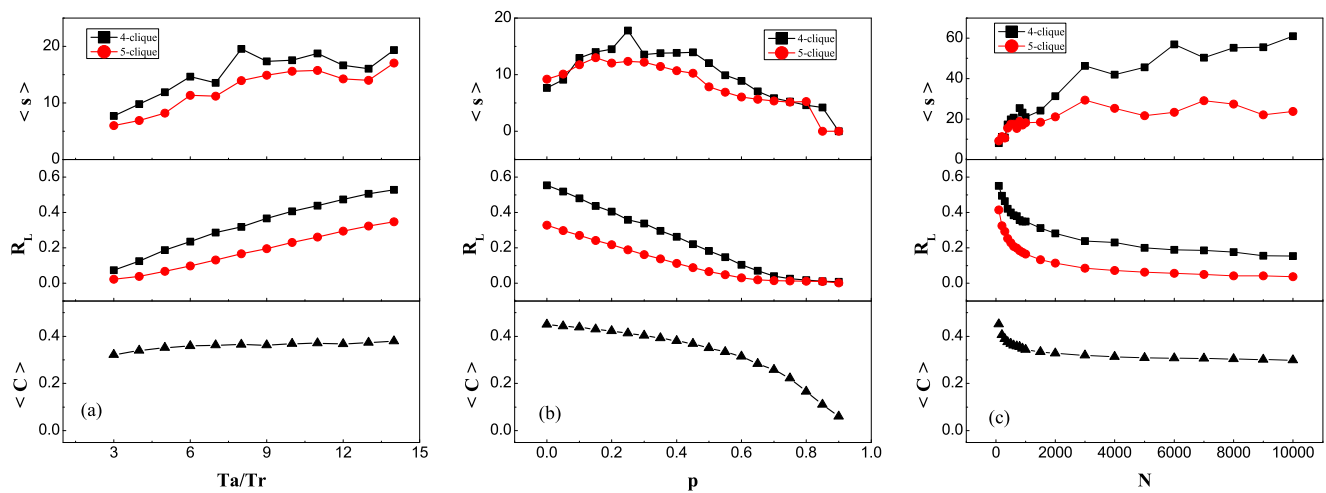}
\caption{ (Color online) Characterization of the influence of
parameters on the properties. Average clustering coefficient
$\langle C\rangle$, relative largest community size $R_L$ and
average community size $\langle s\rangle$ as a function of $T_a/T_r$
(a), as a function of $p$ (b) and as a function of $N$. Results are
averaged over $100$ realizations. Other parameters are the same as
those of figure \ref{fig1}. } \label{fig4}
\end{center}
\end{figure*}

In Fig. \ref{fig3}, we particularly provided two examples to
demonstrate how the network structure, such as the degree
distribution and the community size distribution, varies with the
uniform coupling strength $\gamma$. For small coupling strength,
e.g., $\gamma=0.2$, it is found that basically there are no
communities as shown in Fig. \ref{fig2} (a). This is because of the
fact that when the coupling strength is small, the interactions
among nodes are weak. Thus it is difficult to form dynamically
coherent state on the network. In this case, most new links are
generated randomly. As a result, the network is of random nature and
the degree approximately follows an exponential distribution as
shown in Fig. \ref{fig3} (a). Nevertheless, when the coupling
strength becomes large enough, e.g., $\gamma>2$, the formation of
communities is greatly enhanced by the dynamical interactions among
nodes on the network. In this case, we observe approximate power-law
distribution of degree as shown in Fig. \ref{fig3} (a) in a certain
parameter regime. In Fig. \ref{fig3} (a), it is also seen that the
power-law distribution has the ``droop head'' shape, which has been
observed in many empirical networks \cite{CN:Mobile}. In our study,
it is found that the distribution of community sizes also
approximately follows a power law as shown in Fig. \ref{fig3} (b),
 a phenomenon which has been observed in many empirical networks
\cite{k-clique,communitysize}. In particular, we computed the
average nearest neighbor degree $k_{nn,i}=(1/k_i)\sum_j^Na_{ij}k_j$
to characterize degree-degree correlations in the network
\cite{correlation}. Averaging this quantity over the whole network
gives the averaged nearest neighbor degree $k_{nn}(k)$. As shown in
Fig.\ref{fig3} (c), its distribution turns out to be power-law with
positive exponent. This obviously demonstrates that the formed
network is assortative. Moreover, we also computed the average
clustering coefficient $C(k)$. It decays with the degree $k$ as
shown in Fig. \ref{fig3} (d). This property has been commonly found
in many empirical networks \cite{correlation,clustering3}.

In this work, we have further investigated how the formed network
structures depend on the main control parameters, i.e., $T_a/T_r$,
$p$, and the final network size $N$. The results are presented in
Fig. \ref{fig4}. As shown in Fig. \ref{fig4}(a), the parameter
$T_a/T_r$ has remarkable effect on the properties of the formed
networks. The main network quantities, such as $\langle s \rangle$,
$R_L$, and $C$, all increase with the increase of  $T_a/T_r$. In our
model, the ratio $T_a/T_r$ represents the interaction between two
opposite regulations. $T_a$ is the time period to randomly build a
link, while $T_r$ is the time period to ``purposely" build a link.
The former manipulation is in favor of increasing randomness in the
network, while the latter manipulation helps generate triadic
closure inside the network. Larger  $T_a/T_r$ means that the second
factor is dominant, and this explains why network quantities
 $\langle s \rangle$, $R_L$, and $C$ increase when
 $T_a/T_r$ becomes larger.
 In our model, the parameter $p$ is the probability
deleting links with small similarity distance. When $p$ is
relatively small, this manipulation helps form dynamical groups,
where oscillators within the same group have similar dynamical
states, and this process in turn enhances the formation of
topological communities in the network. However, with further
increase of $p$, according to Eq. (\ref{degree-distribution}), the
network gradually becomes too sparse to form any distinct
communities. Therefore, there exists an optimal parameter range
$0.2<p<0.4$, where communities with various sizes can be formed in
the network. In our model, the final network size $N$ is
proportional to the total integration time. As shown in Fig.
\ref{fig4}(c), all three network quantities $\langle s \rangle$,
$R_L$, and $C$ become approximately stationary when the network
evolves for a long time. Particularly, it is found that the
clustering coefficient $\langle C\rangle$ in the formed networks is
generally large (near $0.3$).
 This property is consistent with empirical observations
\cite{CN:Mobile,clustering3}.

To summarize, we have studied a network-growing model of phase
oscillators, in which the dynamics and the network topology interact
with each other and concurrently evolve. Following simple
node-adding and link-adding/removing rules, the model exhibits
several interesting behaviors. Within a certain parameter range, the
dynamical communities and the topological modules can spontaneously
emerge in the network. It is found that in the formed networks, the
degree and the community size approximately satisfy power-law
distributions. Furthermore, the formed networks show assortative
connection patterns and exhibit high clustering coefficients. Our
study also reveals that the interaction strength among nodes on
network can essentially determine the formation of network
structures, both dynamically and topologically. This is an important
point which has been ignored or over simplified by many previous
network-growing models. The findings in this work capture the
typical properties of many realistic networks. Thus it is helpful
for us to further understand the complicated interactions between
the network topology and the dynamics.

This work is supported by Temasek Laboratories at National
University of Singapore through the DSTA Project No. POD0613356. SGG
is sponsored by the Science and Technology Commission of Shanghai
Municipality under grant no. 10PJ1403300, and also by the NNSF of
China under grant no. 11075056.

\end{document}